\documentclass[12pt,prc,preprint,showpacs]{revtex4}
\usepackage{slashed,graphicx}
\tighten

\begin{document}

\title{Constraining a possible time variation of the gravitational constant $G$ \\
with terrestrial nuclear laboratory data}
\author{Plamen~G.~Krastev and Bao-An Li}
\address{Department of Physics, Texas A\&M University -- Commerce, Commerce, TX 75429, U.S.A}
\date{\today}

\email[P. G. Krastev: ]{Plamen_Krastev@tamu-commerce.edu}

\email[Bao-An Li: ]{Bao-An_Li@tamu-commerce.edu}

\begin{abstract}
Testing the constancy of the gravitational constant $G$ is a
longstanding fundamental question in natural science. As first
suggested by Jofr\'{e}, Reisenegger and
Fern\'{a}ndez~\cite{Jofre:2006ug}, Dirac's hypothesis of a
decreasing gravitational constant $G$ with time due to the expansion
of the Universe would induce changes in the composition of neutron
stars, causing dissipation and internal heating. Eventually, neutron
stars reach their quasi-stationary states where cooling, due to
neutrino and photon emissions, balances the internal heating. The
correlation of surface temperatures and radii of some old neutron
stars may thus carry useful information about the rate of change of
$G$. Using the density dependence of the nuclear symmetry energy,
constrained by recent terrestrial laboratory data on isospin
diffusion in heavy-ion reactions at intermediate energies, and the
size of neutron skin in $^{208}Pb$, within the {\it gravitochemical
heating} formalism developed by Jofr\'{e} et
al.~\cite{Jofre:2006ug}, we obtain an upper limit for the relative
time variation $|\dot{G}/G|$ in the range $(4.5-21)\times
10^{-12}yr^{-1}$.
\end{abstract}
 \maketitle

%\begin{multicols}{2}
%\renewcommand{\thesection}{\arabic{section}}
\section{Introduction}
The question whether or not the fundamental constants of nature vary
with time has been of considerable interest in physics. The
constancy of the gravitational coupling parameter $G$ was first
addressed in 1937 by Dirac~\cite{Dirac:1937ti} who suggested that
the gravitational force might be weakening due to the expansion of
the universe. Although general relativity assumes a strictly
constant $G$, time variations of the Newton's constant are predicted
by some alternative theories of gravity~\cite{BD1961} and a number
of modern cosmological models~\cite{Krauss:1995yb,Bonanno:2001hi}.
Many theoretical approaches, such as models with extra
dimensions~\cite{Overduin:1998pn}, string
theories~\cite{Horava:1995qa,Damour:2002mi,Damour:2002nv}, and
scalar-tensor quintessence models
\cite{Zlatev:1998tr,Armendariz-Picon:2000dh,Armendariz-Picon:2000ah,Steinhardt:1999nw,Hebecker:2000au,Hebecker:2000zb},
have been proposed in which the gravitational coupling parameter
becomes a time-dependent quantity.  Nowadays the debate over the
constancy of $G$ has been revived by recent astronomical
observations~\cite{Perlmutter:1998np,Riess:2000yp} of distant
high-red-shift type Ia supernovae suggesting that presently the
Universe is in a state of accelerated
expansion~\cite{Bonanno:2001hi}. This acceleration can be
interpreted in terms of a ``dark energy'' with negative pressure, or
alternatively by allowing a time variation of the gravitational
constant~\cite{Garcia-Berro:2005yw}. Soon after Dirac had published
his hypothesis~\cite{Dirac:1937ti}, Chandrasekhar~\cite{chandra} and
Kothari~\cite{Kothari} pointed out that a decreasing $G$ with time
could have some detectable astrophysical consequences. Since then
many attempts have been made to find astrophysical signs due to the
possible time variation of $G$. However, there is no firm conclusion
so far (see Ref.~\cite{Uzan:2002vq} for a review). Interestingly
though, as pointed out by Uzan~\cite{Uzan:2002vq}, contrary to most
of the other fundamental constants, as the precision of the
measurements increased the discrepancy among the measured values of
$G$ also increased. This circumstance led $CODATA$ (Committee on
Data for Science and Technology) to raise the relative uncertainty
for $G$~\cite{Uzan:2002vq} by a factor of about 12 in 1998. Some of
the previous upper limits on the time variation of $G$, as obtained
by different experiments/methods, are summarized in Table 1 (adapted
from Reisenegger et al.~\cite{RFJ}). Given the current status of
both theory and experiment, it is fair to say that whether or not
the gravitational constant varies with time is still an open
question and therefore additional work is necessary to investigate
further this fundamental issue.

\renewcommand{\thetable}{\arabic{table}}
\begin{table}[!h]
\centering \caption{Upper bounds on the time variations of $G$
(adapted from Reisenegger et al.~\cite{RFJ}). The first column lists
the method used. The second column contains the upper limit on the
time variation of $G$, most usefully expressed as $|\dot{G}/G|$, the
third column is a rough time scale over which each experiment is
averaging this variation, and the last column is the corresponding
reference. The first two experiments on the list probe the variation
of $G$ from the early Universe to the present time. The next four
experiments are sensitive to long time-scales, but without reaching
to the early Universe. And the last three experiments on the list
probe the change of $G$ over short time-scales of years and
decades.} \vspace{5mm}
\begin{tabular}{cccc}
\hline Method  & $|\dot{G}/G|_{max}[10^{-12}yr^{-1}]$ & Time scale
[yr] & Reference\\
\hline\hline
Big Bang Nucleosynthesis     & 0.4 & $1.4\times 10^{10}$ & \cite{CDK2004}     \\
Microwave Background         & 0.7 & $1.4\times 10^{10}$ & \cite{NCS2004}     \\
Gloubular Cluster Isochrones & 35  & $10^{10}$           & \cite{Degl1996}    \\
Binary Neutron Star Masses   & 2.6 & $10^{10}$           & \cite{Thorsett1996}\\
Helioseismology              & 1.6 & $4\times 10^{9}$    & \cite{GKD1998}     \\
Paleontology                 & 20  & $4\times 10^{9}$    & \cite{ER1977}      \\
Lunar Laser Ranging          & 1.3 & 24                  & \cite{WTB2004}     \\
Binary Pulsar Orbits         & 9   & 8                   & \cite{KTR1994}     \\
White Dwarf Oscillations     & 250 & 25                  & \cite{BGI2002}     \\
\hline
\end{tabular}
\end{table}

Recently a new method, called {\it gravitochemical
heating}~\cite{Jofre:2006ug}, has been introduced to constrain a
hypothetical time variation in $G$, most frequently expressed as
$|\dot{G}/G|$. In Ref.~\cite{Jofre:2006ug} the authors suggested
that such a variation of the gravitational constant would perturb
the internal composition of a neutron star, producing entropy which
is partially released through neutrino emission, while a similar
fraction is eventually radiated as thermal photons. A constraint on
the time variation of $G$ is achieved via a comparison of the
predicted surface temperature with the available empirical value of
an old neutron star~\cite{Kargaltsev:2003eb}. The gravitochemical
heating formalism is based on the results of Fern\'{a}ndez and
Reisenegger~\cite{Fernandez:2005cg} (see
also~\cite{Reisenegger:1994be}) who demonstrated that internal
heating could result from spin-down compression in a rotating
neutron star ({\it rotochemical heating}). In both cases (gravito-
and rotochemical heatings) predictions rely heavily on the equation
of state (EOS) of stellar matter used to calculate the neutron star
structure. Accordingly, detailed knowledge of the EOS is critical
for setting a reliable constraint on the time variation of $G$. The
global properties of neutron stars such as masses, radii, moments of
inertia, thermal evolution, etc have been studied extensively, see,
e.g.,
\cite{Lattimer:2000kb,Lattimer:2004pg,Prakash:2001rx,Yakovlev:2004iq,Heiselberg:2000dn,Heiselberg:1999mq,Steiner:2004fi,KS1}.
Generally, predictions differ widely, mainly, due to the
uncertainties of the equations of state employed in neutron star
structure calculations~\cite{KS1}. Therefore, determining the EOS of
stellar matter is a question of central importance with an answer
requiring understanding of proper nuclear physics.

Using the well tested parabolic approximation the EOS of isospin
asymmetric nuclear matter is written as
\begin{equation}
e(\rho,\alpha)=e(\rho,0)+e_{sym}(\rho)\alpha^2,
\end{equation}
which manifests the separation of the EOS into isospin symmetric
(the energy per particle of symmetric nuclear matter) and isospin
asymmetric contributions (the nuclear symmetry energy $e_{sym}$). In
the above expression $\alpha=(\rho_n-\rho_p)/\rho$ is the usual
asymmetry parameter, $\rho=\rho_n+\rho_p$ is the baryon number
density, and $\rho_n$ and $\rho_p$ are the neutron and proton
densities respectively. While the EOS of symmetric nuclear matter
$(\alpha=0)$ is relatively  well understood, the density dependence
of the symmetry energy, $e_{sym}$, is still very poorly constrained
especially at high densities. Variations in the $e_{sym}$ predicted
by various models often yield dramatically different predictions for
properties of neutron stars (e.g., see Ref.~\cite{KS1}). Because of
its importance for neutron star structure, determining the density
dependence of the symmetry energy has been a high-priority goal for
the intermediate energy heavy-ion community. Although extracting the
symmetry energy is not an easy task due to the complicated role of
isospin in reaction dynamics, several promising probes of the
symmetry energy have been
suggested~\cite{LKR1997,BALi2000,BALi2002,LKB1998} (see also
Refs.~\cite{LS2001,Danielewicz:2002pu,Baran:2004ih} for reviews).

Some significant progress has been made recently in determining the
density dependence of $e_{sym}$ using: (1) isospin diffusion in
heavy-ion reactions at intermediate energies as a probe of both the
magnitude and slope of the symmetry energy around the saturation
density ($\rho_0\approx
0.16fm^{-3}$)~\cite{Shi:2003np,Tsang,Chen:2004si,Steiner:2005rd,Li:2005jy,Chen:2005ti},
(2) flow in heavy-ion collisions at higher energies to constrain the
equation of state of symmetric nuclear
matter~\cite{Danielewicz:2002pu}, and (3) the sizes of neutron skins
in heavy nuclei to constrain $e_{sym}(\rho)$ at subsaturation
densities~\cite{Steiner:2004fi,Horowitz:2000xj,Horowitz:2002mb,RP2005}.

In this work, we combine recently obtained diffusion data,
information from flow observables, studies on the neutron skin of
$^{208}Pb$, and other information to constrain a possible
time-variation of the gravitational constant $G$ through the
gravitochemical formalism~\cite{Jofre:2006ug}. We do not aim to add
anything fundamental to the original method of
Ref.~\cite{Jofre:2006ug}. Our objective is to provide a restrictive
upper limit for the time variation of $G$, applying an EOS
constrained by terrestrial empirical data from nuclear reactions
induced by neutron-rich nuclei. After the introductory notes in this
section, we discuss, in some details, the general formalism of the
gravitochemical method. The equations of state used in this study
are outlined briefly in Section III. Our results for the upper
limits of $|\dot{G}/G|$ are presented and discussed in Section IV.
The effects of ``exotic'' (hyperonic/quark) phases in neutron star
matter on a possible time variation of $G$ are addressed in Section
V. We conclude in Section VI with a short summary.

\section{Gravitochemical heating: Formalism}

To provide the reader with a self-contained manuscript, in this
section we recall the main steps leading to the calculation of the
surface temperature of an old neutron star via the gravitochemical
heating method. For a detailed discussion see
Refs.~\cite{Jofre:2006ug,Fernandez:2005cg}. (Conventions and
notation as in the above references.) The simplest neutron star
models assume a composition of nucleons and light leptons, electrons
and muons, which can transform into each other through direct and
inverse $\beta$-reactions. The neutrinos ($\nu$) and antineutrinos
($\bar{\nu}$) produced in these reactions leave the star without
further interactions, contributing to its cooling. In
$\beta$-equilibrium the balance between the rates of direct and
inverse processes is reflected through the following relation among
the chemical potentials of the particle species
\begin{equation}
\mu_n-\mu_p=\mu_e=\mu_{\mu}
\end{equation}
As pointed out in Ref.~\cite{Jofre:2006ug} a time-variation of $G$
would cause continuously a perturbation in the stellar density and
since the chemical potentials are density-dependent the star thus
always departs from $\beta$-equilibrium. This departure is
quantified by the chemical imbalances
\begin{eqnarray}
\eta_{npe}&=&\delta\mu_n-\delta\mu_p-\delta\mu_e \\
\eta_{np\mu}&=&\delta\mu_n-\delta\mu_p-\delta\mu_{\mu}
\end{eqnarray}
where $\delta\mu_i=\mu_i-\mu_i^{eq}$ is the deviation of the
chemical potential of particle species $i$ ($i=n,p,e,\mu$) from its
equilibrium value at a given pressure. The chemical imbalances
enhance the rates of reactions driving the star to a new equilibrium
state. If $G$ changes continuously with time the star will be always
out of equilibrium, storing an excess of energy that is dissipated
as internal heating and enhanced neutrino
emission~\cite{Jofre:2006ug}.

The evolution of the internal temperature is given by the thermal
balance equation
\begin{equation}
\dot{T}^{\infty}=\frac{1}{C}[L_H^{\infty}-L_{\nu}^{\infty}-L_{\gamma}^{\infty}]
\end{equation}
where $C$ is the total heat capacity of the star, $L_H^{\infty}$ is
the total power released by heating mechanisms, $L_{\nu}^{\infty}$
is the total neutrino luminosity, and $L_{\gamma}^{\infty}$ is the
photon luminosity (``$\infty''$ labels the quantities as measured by
a distant observer). The evolution of the red-shifted chemical
imbalances is governed by
\begin{eqnarray}
\dot{\eta}_{npe}^{\infty}&=&\delta\dot{\mu}_n^{\infty}-\delta\dot{\mu}_p^{\infty}
-\delta\dot{\mu}_e^{\infty}\\
\dot{\eta}_{np\mu}^{\infty}&=&\delta\dot{\mu}_n^{\infty}-\delta\dot{\mu}_p^{\infty}
-\delta\dot{\mu}_{\mu}^{\infty}
\end{eqnarray}
These equations can be written as~\cite{Jofre:2006ug}
\begin{eqnarray}
\dot{\eta}_{npe}^{\infty}=&-&[A_{D,e}(\eta_{npe}^{\infty},T^{\infty})+A_{M,e}(\eta_{npe}^{\infty},T^{\infty})]
-[B_{D,e}(\eta_{np\mu}^{\infty},T^{\infty})+B_{M,e}(\eta_{np\mu}^{\infty},T^{\infty})]\\
\dot{\eta}_{np\mu}^{\infty}=&-&[A_{D,\mu}(\eta_{npe}^{\infty},T^{\infty})+A_{M,\mu}(\eta_{npe}^{\infty},T^{\infty})]
-[B_{D,\mu}(\eta_{np\mu}^{\infty},T^{\infty})+B_{M,\mu}(\eta_{np\mu}^{\infty},T^{\infty})]
\end{eqnarray}
The functions $A$ and $B$ quantify the effect of reactions toward
restoring chemical equilibrium, and thus have the same sign as
$\eta_{npl}$ ($l=e,\mu$)~\cite{Fernandez:2005cg}. The subscripts $D$
refers to the so-called direct Urca cooling processes
\begin{eqnarray}
n&\rightarrow&p+l+\bar{\nu}\nonumber\\
p+l&\rightarrow&n+\nu,
\end{eqnarray}
which are fast but possibly forbidden by the energy-momentum
conservations when the proton fraction is
low~\cite{Yakovlev:2004iq}. The subscripts $M$ refers to the
modified Urca reactions
\begin{eqnarray}
n+N&\rightarrow&p+N+l+\bar{\nu}\nonumber\\
p+l+N&\rightarrow&n+N+\nu,
\end{eqnarray}
which are slow and an additional nucleon or nucleus $N$ must
participate in order to conserve momentum~\cite{Yakovlev:2004iq}.
The constants $C_{npe}$ and $C_{np\mu}$ quantify the departure from
chemical equilibrium due to a time-variation of $G$. They can be
written as~\cite{Jofre:2006ug}
\begin{eqnarray}
C_{npe}&=&(Z_{npe}-Z_{np})I_{G,e}+Z_{np}I_{G,p}\nonumber\\
C_{np\mu}&=&(Z_{np\mu}-Z_{np})I_{G,\mu}+Z_{np}I_{G,p}
\end{eqnarray}
Here $I_{G,i}=(\partial N_i^{eq}/\partial G)_A$ is the change of the
equilibrium number of particle species $i$ ($i=n,p,e,\mu$),
$N_i^{eq}$, due to the variation of $G$ and $Z$ are constants
depending only on the stellar structure~\cite{Jofre:2006ug}.
Equations (5), (8), and (9) determine completely the thermal
evolution of a neutron star with gravitochemical heating. The main
consequence of this mechanism is that eventually the star arrives at
a quasi-equilibrium state, with heating and cooling balancing each
other~\cite{Jofre:2006ug}. The properties of this stationary state
can be obtained by solving simultaneously Eqs.~(5), (8), and (9) by
setting
$\dot{T}^{\infty}=\dot{\eta}_{npe}^{\infty}=\dot{\eta}_{np\mu}^{\infty}=0$.
The existence of a quasi-equilibrium state makes it possible, for a
given value of $|\dot{G}/G|$, to compute the temperature of an old
neutron star without knowing its exact age~\cite{Jofre:2006ug},
since, due to the independence of the solution from the initial
conditions, it is unnecessary to model the complete evolution of the
chemical imbalances and temperature.

If only the modified Urca reactions are allowed, for a given stellar
model, it is possible to derive an analytic expression relating the
photon luminosity in the stationary state, $L_{\gamma,eq}^{\infty}$,
to $|\dot{G}/G|$. This is because the longer time scale required to
reach a stationary state, when only modified Urca processes operate,
results in chemical imbalances satisfying
$\eta_{npl}>>k_BT$~\cite{Jofre:2006ug}. Under these conditions the
photon luminosity in the quasi-equilibrium state is given by
%\begin{widetext}
\begin{equation}
L_{\gamma,eq}^{\infty}=C_M\left(\frac{k_BG}{C_H}\right)^{8/7}
\left[\left(\frac{I_{G,e}^8}{\tilde{L}_{Me}}\right)^{1/7}+
\left(\frac{I_{G,\mu}^8}{\tilde{L}_{M\mu}}\right)^{1/7}\right]
\left|\frac{\dot{G}}{G}\right|^{8/7}
\end{equation}
%\end{widetext}
The meaning of the constants $C_M$ and $C_H$, and the functions
$\tilde{L}_{M_i}$ ($i=e,\mu$) is explained in
Refs.~\cite{Fernandez:2005cg,Jofre:2006ug}. From
$L_{\gamma,eq}^{\infty}$ the neutron-star surface temperature can be
calculated by assuming an isotropic blackbody spectrum
\begin{equation}
L_{\gamma,eq}^{\infty}=4\pi\sigma R^2_{\infty}(T_s^{\infty})^4
\end{equation}
with $\sigma$ the Stefan-Boltzmann constant and $R_{\infty}$ the
red-shifted radius of the star. In the case when only slow
$\beta$-reactions operate, we write the stationary surface
temperature as
\begin{equation}
T_s^{\infty}=\tilde{\cal D}\left|\frac{\dot{G}}{G}\right|^{2/7},
\end{equation}
where the function $\tilde{\cal D}$ is a quantity depending only on
the stellar model and the equation of state. On the other hand, if
at higher densities the proton fraction is large enough so that the
direct Urca reactions are allowed, the thermal evolution of a
neutron star with gravitochemical heating needs to be modeled by
solving numerically the coupled Eqs. (5), (8) and (9).

As demonstrated by Jofr\'{e} et al.~\cite{Jofre:2006ug} the
formalism outlined here can be applied to constrain the value of
$|\dot{G}/G|$, provided one knows (i) the surface temperature of a
neutron star, and (ii) that the star is certainly older than the
time-scale necessary to reach a quasi-stationary state. So far the
only object satisfying both conditions is  PSR J0437-4715, which is
the closest millisecond pulsar to our solar system. Its surface
temperature was deduced from ultraviolet
observations~\cite{Kargaltsev:2003eb} while its mass was determined
by Hotan et al.~\cite{HBO2006} to be in the range
$M_{PSR}=(1.1-1.5)M_{\odot}$. (Another mass constraint,
$M_{PSR}=1.58\pm 0.18M_{\odot}$, was given previously by van Straten
et al.~\cite{vanStraten:2001zk}.) To constrain the value of
$|\dot{G}/G|$ one, therefore, needs to consider neutron-star models
in the above mass range and calculate the surface temperature for
each stellar configuration.

\section{Equation of State and Neutron Star Structure}
Clearly, predictions of the surface temperature and, in turn, value
of $|\dot{G}/G|$ depend heavily on the EOS of neutron-star matter
since the later is critical for determining the neutron-star
structure. Currently, theoretical predictions of the EOS of
neutron-rich matter diverge widely mainly due to the uncertain
density dependence of the nuclear symmetry energy. Consequently, to
provide a stringent constraint on the time variation of $G$, one
should attempt to reduce the uncertainty due to the $e_{sym}(\rho)$.
Recently available nuclear reaction data allowed us to constrain
significantly the density dependence of the symmetry energy mostly
in the sub-saturation density region. While high energy radioactive
beam facilities under construction will provide a great opportunity
to pin down the high density behavior of the nuclear symmetry energy
in the future. In this work, we apply the gravitochemical method
with several EOSs describing matter of purely nucleonic ($npe\mu$)
as wells as hyperonic and hybrid stars. Among the nucleonic matter
EOSs, we pay special attention to the one calculated with the MDI
interaction~\cite{Das:2002fr}. The symmetry energy $e_{sym}(\rho)$
of the MDI EOS is constrained in the sub-saturation density region
by the available nuclear laboratory data, while in the high-density
region we assume a continuous density functional. The EOS of
symmetric matter $e(\rho,0)$ for the MDI interaction is constrained
up to about five times the normal nuclear matter density by the
available data on collective flow in relativistic heavy-ion
reactions.

Let us first briefly recall here the main ingredients of the MDI EOS
following Ref.~\cite{Li:2005sr}. The MDI EOS corresponds to the
single-particle potential
\begin{eqnarray}
U(\rho,\alpha,\vec{p},\tau,x)&=&A_u(x)\frac{\rho_{\tau'}}{\rho_0}+A_l(x)\frac{\rho_{\tau}}{\rho_0}
+B\left(\frac{\rho}{\rho_0}\right)^{\sigma}(1-x\alpha^2)-8\tau
x\frac{B}{\sigma+1}\frac{\rho^{\sigma-1}}{\rho_0^{\sigma}}\alpha\rho_{\tau'}\nonumber\\
&+&\frac{2C_{\tau,\tau}}{\rho}\int
d^3p'\frac{f_{\tau}(\vec{r},\vec{p}')}{1+(\vec{p}-\vec{p}')^2/\Lambda}+\frac{2C_{\tau,\tau'}}{\rho}\int
d^3p'\frac{f_{\tau'}(\vec{r},\vec{p}')}{1+(\vec{p}-\vec{p}')^2/\Lambda},
\end{eqnarray}
deduced~\cite{Das:2002fr} from the Gogny interaction. In the above
equation $x$ is a parameter introduced to reflect the largely
uncertain density dependence of the $e_{sym}(\rho)$ as predicted by
various many-body approaches; $\tau(\tau')$ is $1/2$ ($-1/2$) for
neutrons (protons) with $\tau\neq\tau'$; $\sigma=4/3$,
$f_{\tau}(\vec{r},\vec{p})$ is the space distribution function at
coordinate $\vec{r}$ and momentum $\vec{p}$; $A_u$, $A_l$, $B$,
$C_{\tau,\tau}$, $C_{\tau',\tau'}$ and $\Lambda$ are parameters
fixed by fitting the momentum dependence of
$U(\rho,\alpha,\vec{p},\tau,x)$, as predicted by the
Gogny/Hartree-Fock and/or Brueckner-Hartree-Fock (BHF) calculations,
so that the saturation properties of nuclear matter and the value of
the symmetry energy at the saturation density
($e_{sym}(\rho_0)\approx 32 MeV$) are predicted correctly. The
compression modulus of saturated nuclear matter, $\kappa$, is set to
$211MeV$ consistent withe empirical range recently suggested by
Garg~\cite{Garg:2006vc}. More specifically, $B=106.35MeV$ and
$\Lambda=k_F^0$ is the nucleon Fermi momentum in symmetric nuclear
matter. The quantities $A_u(x)$ and $A_l(x)$ depend on the parameter
$x$ according to
\begin{eqnarray}
A_u(x)&=&-95.98-x\frac{2B}{\sigma+1}\\
A_l(x)&=&-120.57+x\frac{2B}{\sigma+1}
\end{eqnarray}

\begin{figure}[!t]
\centering
\includegraphics[totalheight=3.7in]{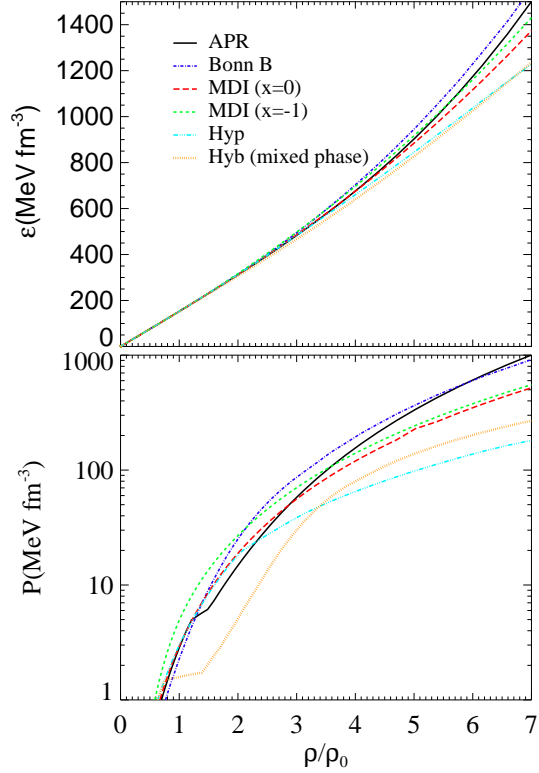}
\vspace{5mm}\caption{(Color online) Equation of state of stellar
matter in $\beta$-equilibrium. The upper panel shows the total
energy density and lower panel the pressure as function of the
baryon number density (in units of $\rho_0$).}
\end{figure}

\begin{figure}[!t]
\centering
\includegraphics[totalheight=3.7in]{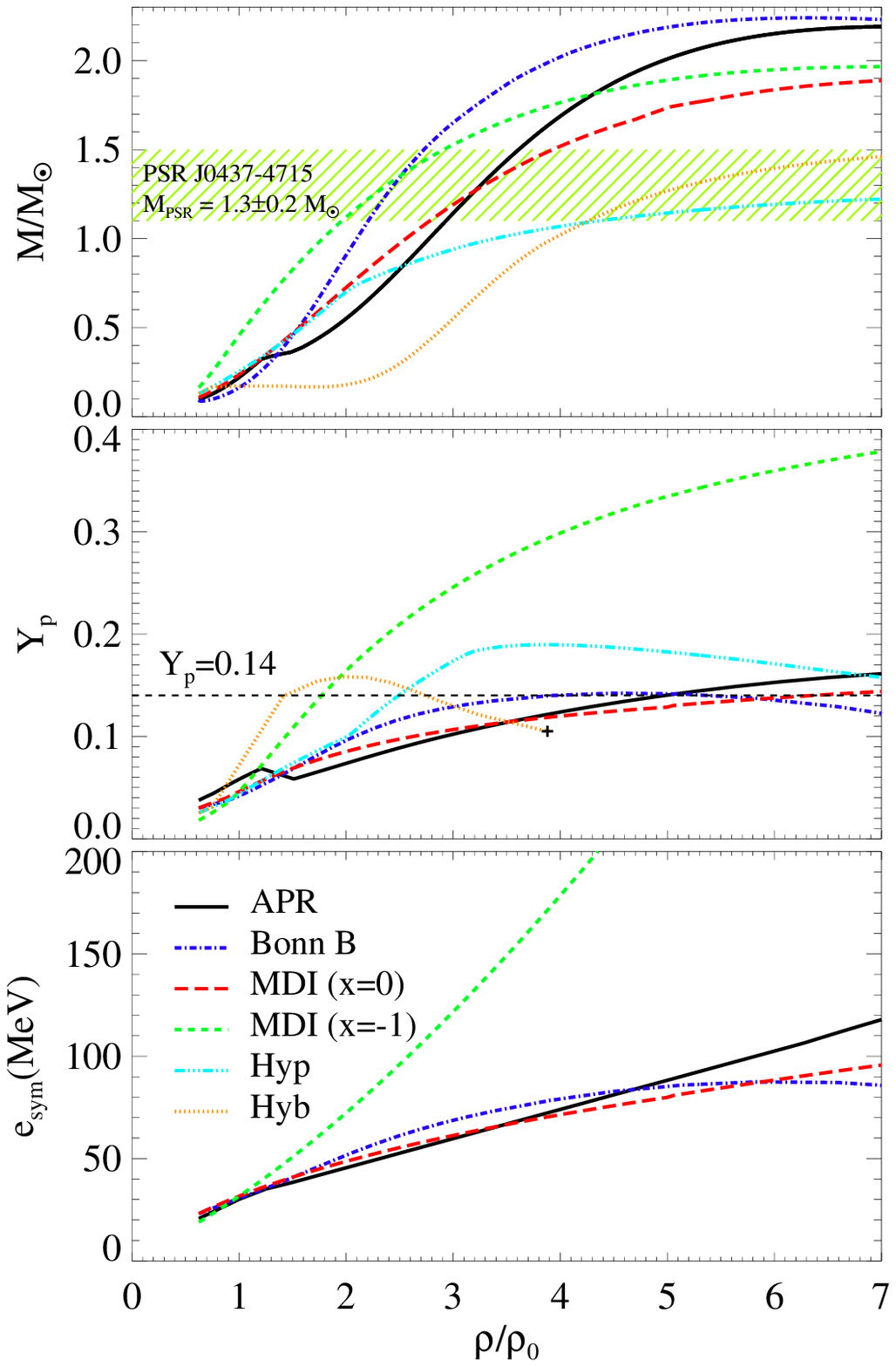}
\vspace{5mm}\caption{(Color online) Neutron star mass, proton
fraction, $Y_p$, and symmetry energy, $e_{sym}$. The upper frame
displays the neutron star mass as a function of baryon number
density. The middle frame shows the proton fraction and the lower
frame the nuclear symmetry energy as a function of density.
(Symmetry energy is shown for the nucleonic EOSs only.) The proton
fraction curve of the Hyb EOS is terminated at the beginning of the
quark phase. The termination point is denoted by a ``cross''
character.}
\end{figure}

The isoscalar potential, evaluated from $U_0=(U_n+U_p)/2$, agrees
very well with predictions from many-body
variational~\cite{Wiringa:1988jt} and recent
Dirac-Brueckner-Hartree-Fock (DBHF)~\cite{Sammarruca:2004cy}
calculations. The underlying EOS has been also successfully tested
against nuclear collective flow data in relativistic heavy-ion
reactions at densities up to $5\rho_0$ $(\rho_0\approx 0.16
fm^{-3})$~\cite{Danielewicz:2002pu,WPKG1988,GWP1990,ZGG1994}. Also
the strength of the symmetry potential estimated from the
single-nucleon potentials via $U_{sym}=(U_n-U_p)/(2\alpha)$ at
$\rho_0$ agrees very well with the Lane potential extracted from
nucleon-nucleus scatterings and $(p,n)$ charge exchange reactions up
to $100$ MeV. The EOS outlined here has been applied recently to
constrain the neutron-star radius~\cite{Li:2005sr} with a suggested
range compatible with the best estimates from observations.

We show the EOSs used in this work in Fig.~1. The upper panel
displays the total energy density, $\epsilon$, as a function of
baryon number density and the lower frame shows total pressure.
Predictions from Akmal et al.~\cite{Akmal:1998cf} with the
$A18+\delta v+UIX^*$ interaction (APR) and recent DBHF calculations
(Bonn B)~\cite{Alonso&Sammarruca1,KS1} with the Bonn B
One-Boson-Exchange (OBE) potential are also shown. In addition to
the pure nucleonic EOSs, we also include one hyperonic (Hyp) and one
hybrid (Hyb) EOSs by the Catania group~\cite{G.F.Burgio}, see e.g.
Ref.~\cite{GFB2007}. The hyperonic EOS shown in Fig.~1 is an updated
version of the one reported in Ref.~\cite{BBS2000}. It has been
calculated within the Brueckner-Hartree-Fock (BHF) approach using
the Argone $v_{18}$ nucleon-nucleon (NN) potential, modified by
nucleon three-body-forces according to the Urbana model, and with
nucleon-hyperon interaction included. The hybrid EOS contains a BHF
hadronic phase ($n,p,e,\mu,\Lambda,\Sigma$) followed by a mixed
phase ($n,p,e,\mu,\Lambda,\Sigma,u,d,s$), and a pure quark phase
($u,d,s,e$), calculated within the MIT bag model (with bag constant
$B=90MeV fm^{-3}$ and $150$ MeV mass of the
s-quark)~\cite{G.F.Burgio}. The hadron-quark phase transition has
been obtained by performing a Gibbs construction~\cite{G.F.Burgio}.
Since, as demonstrated in Refs.~\cite{Li:2005jy,Li:2005sr}, only
equations of state with $x$ between -1 and 0 have symmetry energies
consistent with the isospin diffusion data and measurements of the
skin thickness of $^{208}Rb$, we thus consider only these two
limiting cases. Below the density of approximately $0.07fm^{-3}$ the
equations of state shown in Fig.~1 are supplemented with a crust
EOS~\cite{PRL1995,HP1994} which is more suitable for the low density
regime.

\begin{figure}[!t]
\centering
\includegraphics[totalheight=2.5in]{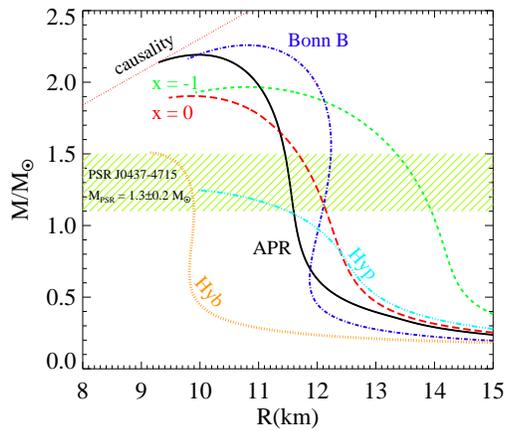}
\caption{(Color online) Mass-radius relation. The shaded region
corresponds to the mass constraint from Hotan et
al.~\cite{HBO2006}.}
\end{figure}

\begin{table}[!t]
\caption{Maximum-mass neutron star models. The first column
identifies the EOS. Remaining columns exhibit the following
quantities for static configurations with maximum mass: total
gravitational mass; radius; central baryon density.} \centering
\vspace{5mm}
\begin{tabular}{lccc}
\hline
EOS                  &  $M_{max}(M_{\odot})$ & $R(km)$ & $\rho_c(fm^{-3})$\\
\hline\hline
MDI(x=0)             & 1.91    &  9.89  & 1.30\\
MDI(x=-1)            & 1.97    & 10.85  & 1.08\\
Bonn B               & 2.24    & 10.88  & 1.08\\
APR                  & 2.19    &  9.98  & 1.14\\
Hyp                  & 1.25    &  9.96  & 1.49\\
Hyb (mixed phase)    & 1.51    &  9.14  & 1.50\\
\hline
\end{tabular}
\end{table}

Fig.~2 displays the neutron star mass (upper panel), the proton
fraction (middle panel) and the nuclear symmetry energy (lower
panel). The shaded region in the upper frame corresponds to the mass
constraint by Hotan et al.~\cite{HBO2006}. From the neutron star
models {\it satisfying this constraint} we observe that the proton
fraction exceeds the direct URCA threshold for predictions from the
$x=-1$ and Hyp EOSs. Here we recall that the fast URCA process
proceeds only for $Y_p$ above $0.14$ due to simultaneous
conservation of energy and momentum~\cite{Lattimer:1991ib}, causing
rapid cooling of neutron stars. Hybrid stars (from the Hyb EOS) cool
rapidly through a quark direct URCA process (see Section V).

The corresponding stellar sequences are shown in Fig.~3. Table 2
summarizes the gravitational masses, radii and central densities of
the {\it maximum-mass} configurations.

\section{Constraining the changing rate of the gravitational constant G}

To constrain the hypothetical time variation of $G$, we consider
stellar models constructed from different equations of state and
calculate the neutron star surface temperature via the
gravitochemical heating method. Since only EOSs with $x$ between -1
and 0 have symmetry energies consistent with terrestrial nuclear
laboratory data~\cite{Li:2005jy,Li:2005sr}, we consider these two
limiting cases as representative of the possible range of neutron
star structures. As demonstrated by Jofr\'{e} et
al.~\cite{Jofre:2006ug}, if one assumes only slow Urca reactions in
the neutron star interior, the stellar photon luminosity and surface
temperature in the stationary state can be evaluated through Eqs.
(13) and (15) respectively. In the most general case, however, when
both direct and modified Urca reactions operate in the neutron star
interior (as for the $x=-1$, $Hyp$ and $Hyb$ EOSs), the analytic
expression (13) becomes a very poor approximation~\cite{RF}.
Therefore, under this circumstance we calculate the surface
temperature (and photon luminosity) in the quasi-equilibrium state
by setting Eqs. (5), (8), and (9) to zero and solving them
simultaneously.

\begin{figure}[!b]
\centering
\includegraphics[totalheight=2.7in]{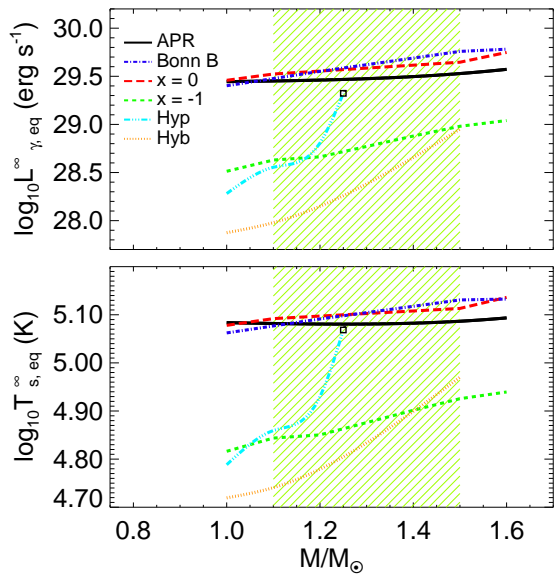}
\vspace{5mm} \caption{(Color online) Stationary photon luminosity
(upper frame) and neutron star surface temperature (lower frame) as
functions of stellar mass, assuming $|\dot{G}/G|=4.5\times
10^{-12}yr^{-1}$. The shaded region corresponds to the mass
constraint form Hotan et al.~\cite{HBO2006}. The open (black) square
character denotes the maximum possible mass attained by the neutron
star models constructed from the hyperonic EOS (Hyp). (The $x=-1$,
$Hyp$ and $Hyb$ EOSs allow for fast cooling in the considered mass
range. For the $Hyb$ EOS, stellar cooling proceeds via a quark
direct URCA process, see Section V.)}
\end{figure}

\begin{figure}[!t]
\centering
\includegraphics[totalheight=2.1in]{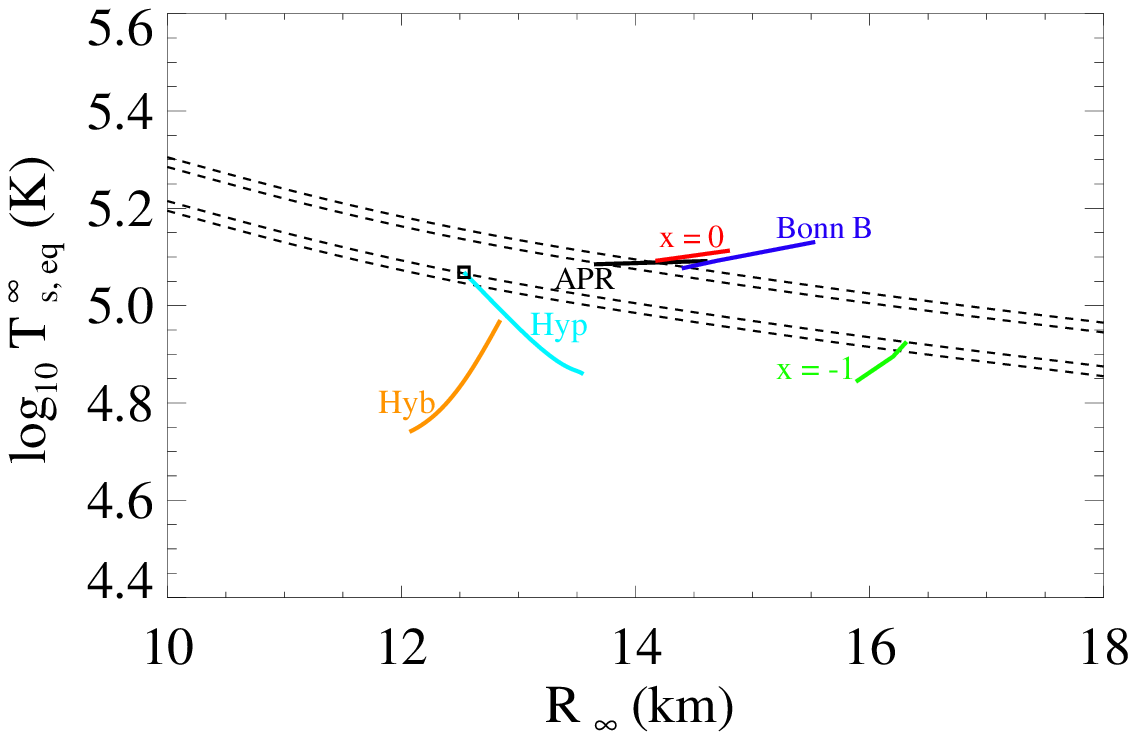}
\vspace{2mm} \caption{(Color online) Neutron star stationary surface
temperature for stellar models satisfying the mass constraint by
Hotan et al.~\cite{HBO2006}. The solid lines are the predictions
versus the stellar radius for the considered neutron star sequences.
Dashed lines correspond to the 68\% and 90\% confidence contours of
the black-body fit of Kargaltsev et al.~\cite{Kargaltsev:2003eb}.
The value of $|\dot{G}/G|=4.5\times 10^{-12}yr^{-1}$ is chosen so
that predictions from the $x=0$ EOS are just above the observational
constraints.}
\end{figure}

In Fig.~4 we show the neutron star stationary photon luminosity
(upper panel) and steady surface temperature (lower panel) versus
stellar mass, assuming $|\dot{G}/G|=4.5\times 10^{-12}yr^{-1}$. The
value of $\dot{G}$ is chosen so that predictions from the $x=0$ EOS
are just above the 90\% confidence contour of Kargaltsev et
al.~\cite{Kargaltsev:2003eb}, see Fig.~5. This upper limit is
consistent with the one by Jofr\'{e} at al.~\cite{Jofre:2006ug}
under the same assumptions. Here we reiterate that we apply the mass
constraint by Hotan et al.~\cite{HBO2006} instead the one by van
Straten et al.~\cite{vanStraten:2001zk} used in
Ref.~\cite{Jofre:2006ug}. (Both mass measurements partially overlap
in the range $(1.4-1.5)M_{\odot}$.) We notice that predictions from
the $x=0$, APR and Bonn B EOSs all lie just above the observational
constraints, with those from the $x=0$ and APR EOSs being very
similar to each other. This observation has been already made in a
previous work~\cite{Li:2005sr} in conjunction with a study of the
neutron star radius and was interpreted in terms of the good
agreement between the corresponding symmetry energies up to about
$5\rho_0$ (see also Fig.~2, lower frame).

\begin{figure}[!t]
\centering
\includegraphics[totalheight=2.7in]{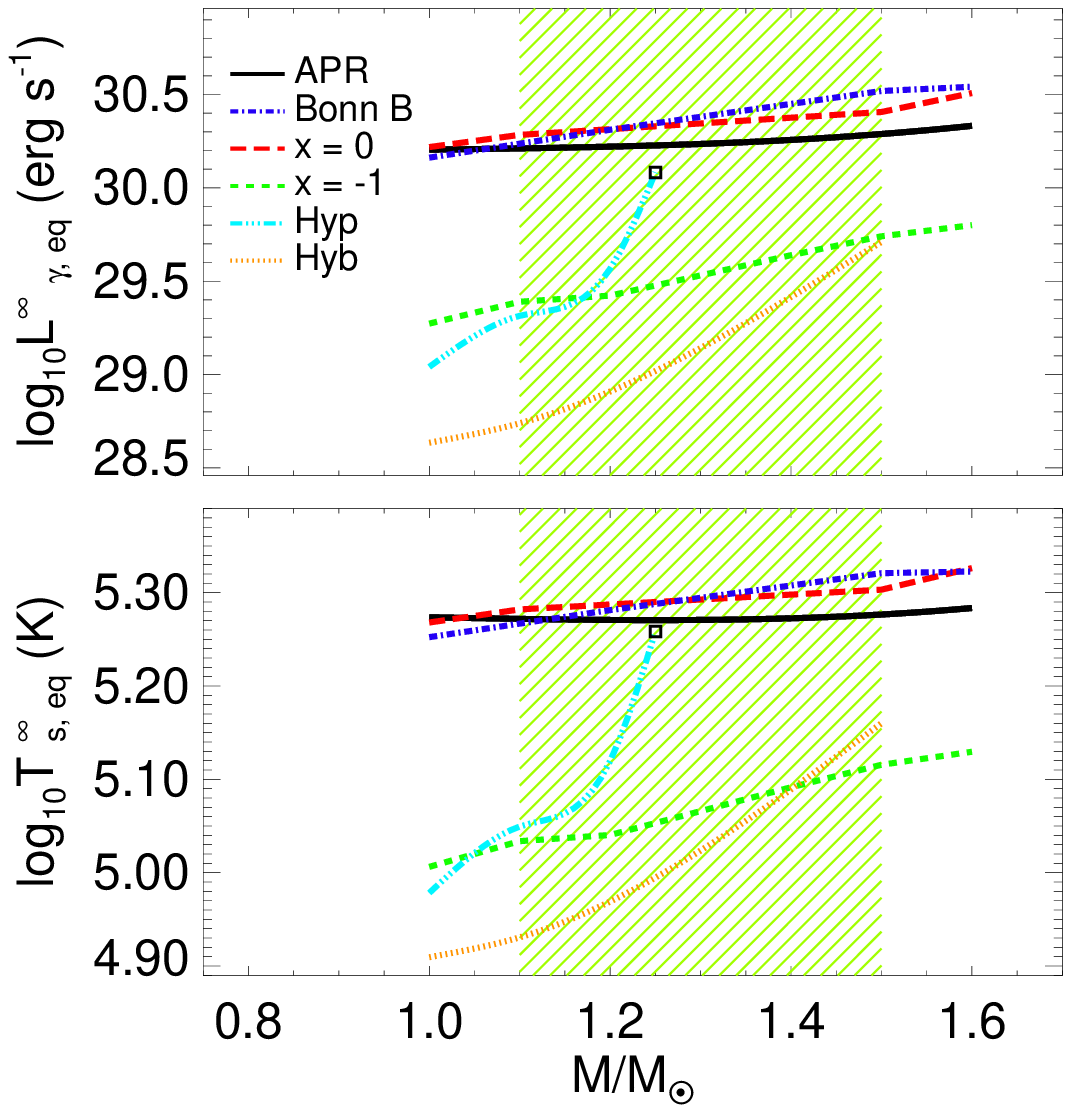}
\vspace{5mm} \caption{(Color online) Same as Fig.~4 but now the
value of $|\dot{G}/G|=2.1\times 10^{-11}yr^{-1}$ is chosen so that
predictions from the x=-1 EOS are just above the observational
constraints. (The $x=-1$ EOS allows direct Urca reactions in the
considered mass range.)}
\end{figure}

\begin{figure}[!t]
\centering
\includegraphics[totalheight=2.1in]{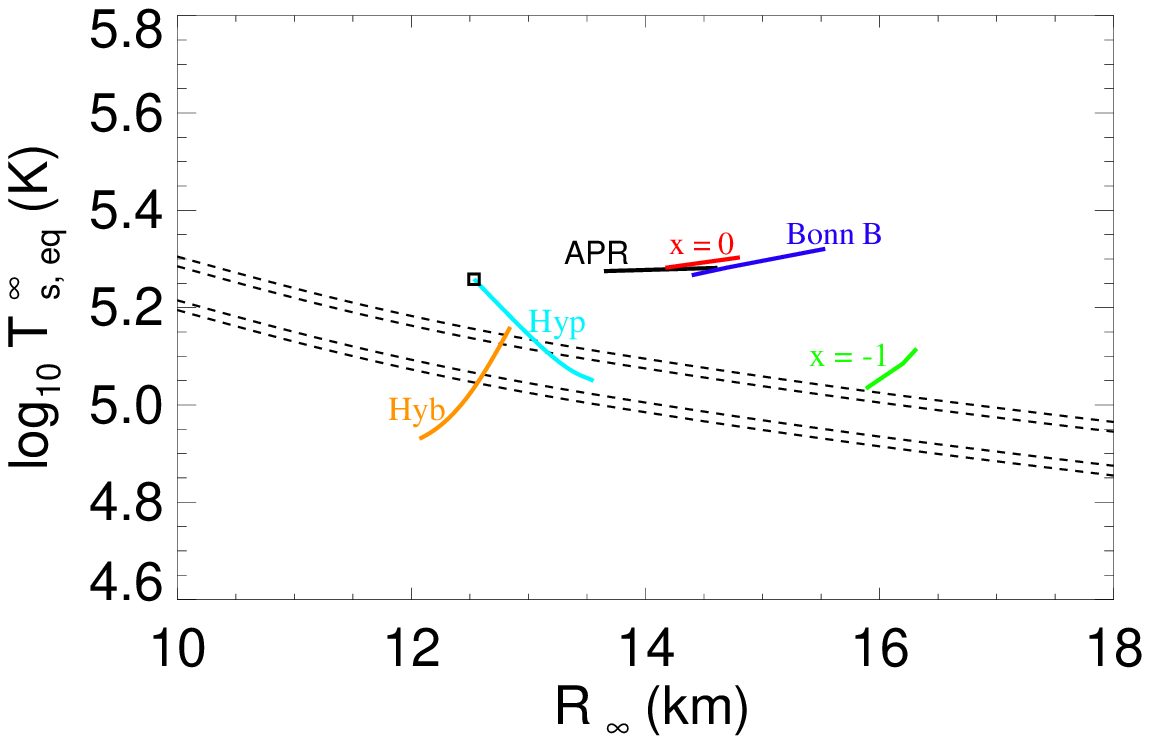}
\vspace{2mm} \caption{(Color online) Same as Fig.~5 but assuming
$|\dot{G}/G|=2.1\times 10^{-11}yr^{-1}$.}
\end{figure}

Figs.~(6) and (7) display predictions assuming
$|\dot{G}/G|=2.1\times 10^{-11}yr^{-1}$. In this case the value of
$\dot{G}$ is chosen so that predictions from the $x=-1$ EOS are just
above the observational constraints. The surface temperature
calculated with the $x=-1$ EOS is noticeably lower than predictions
from EOSs allowing only modified Urca processes. This is due to
opening of the fast Urca channel (see, Fig.~2, lower panel). When
the neutron star mass becomes large enough for the central density
to exceed the direct Urca threshold, the surface temperature drops
because of the faster relaxation toward chemical
equilibrium~\cite{Jofre:2006ug}. The upper limit we provide here is
more restrictive than the one reported by Jofr\'{e} et
al.~\cite{Jofre:2006ug} in the case of fast cooling.

\begin{figure}[!t]
\centering
\includegraphics[totalheight=2.1in]{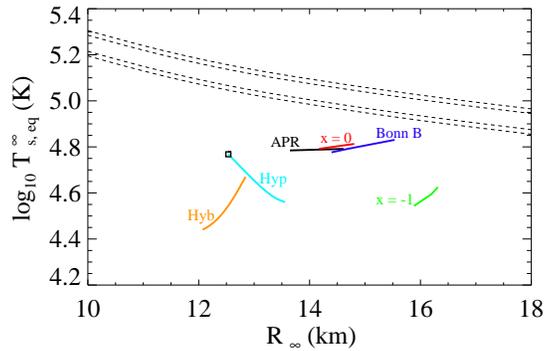}
\vspace{2mm} \caption{(Color online) Same as Fig.~5 but assuming
$|\dot{G}/G|=4\times 10^{-13}yr^{-1}$ after Copi et
al.~\cite{CDK2004}.}
\end{figure}

Before we close the discussion in this section we also compare our
results with recent predictions from big bang nucleosynthesis
(BBN)~\cite{CDK2004}. The stationary surface temperature in Fig.~8
is calculated assuming $|\dot{G}/G|=4\times 10^{-13}yr^{-1}$ after
Copi et al.~\cite{CDK2004}. We observe that in this case the
resulting surface temperatures are below the observational contour
of Kargaltsev et al.~\cite{Kargaltsev:2003eb}. There are several
possible reasons which could account for this discrepancy: (i) The
time scales of both methods are quite different. While the approach
applied by Copi et al.~\cite{CDK2004} probes the constancy of $G$
over the age of Universe (see Table~1), the gravitochemical heating
method is sensitive to variations in $G$ over shorter
periods~\cite{Jofre:2006ug,RFJ}, and falls closest to the second
group of methods listed in Table~1. (ii) We do not consider the
rotochemical heating~\cite{Fernandez:2005cg} in this paper. On the
other hand, the rotochemical heating, which is believed to play an
important role in the thermal evolution of millisecond pulsars,
cannot explain alone the observations of Kargaltsev et
al.~\cite{Kargaltsev:2003eb}. Therefore, additional heating
mechanisms need to be considered~\cite{Fernandez:2005cg}. In fact,
if taken together the roto- and gravitochemical heatings could, in
principle, explain the empirical surface temperature of PSR
J0437-4715, where a smaller upper limit for $|\dot{G}/G|$ should be
assumed~\cite{RFJ}. Under this circumstance, our predictions and
those of Copi et al.~\cite{CDK2004} would be relatively close. (iii)
Superfluidity is not taken into account. As pointed out in
Refs.~\cite{Fernandez:2005cg,Jofre:2006ug,RFJ}, this would raise the
surface temperature due to lengthening the time-scales needed to
reach equilibrium.

\section{Effects of Hyperons and Quarks}

Although the major goal of our work is to constrain the changing
rate of $G$ by applying an EOS with constrained symmetry energy, our
analysis would be incomplete without a discussion of the possible
effects of ``exotic'' states of matter in neutron stars. Namely, in
what follows we discuss the effects of hyperons and quarks.

{\it Hyperons.-} While at normal nuclear matter densities,
$\rho\approx\rho_0$, neutron star matter consists of only nucleons
and light leptons ($e^-,\mu^-$), at higher densities several other
species of particles are expected to appear due to the rapid rise of
the baryon chemical potentials with density~\cite{BBS2000}. The
first particles to appear after the muons are the $\Sigma^-$ and
$\Lambda^0$ hyperons. Strange baryons appear mainly in reactions
such as~\cite{YKGH2001}
\begin{eqnarray}
n+n&\Longleftrightarrow&\Sigma^-+p\\
n+n&\Longleftrightarrow&n+\Lambda^0\\
n+\Lambda^0&\Longleftrightarrow&\Sigma^-+p
\end{eqnarray}
The equilibrium conditions for these reactions read
\begin{eqnarray}
2\mu_n&=&\mu_{\Sigma^-}+\mu_p\\
\mu_n&=&\mu_{\Lambda^0}
\end{eqnarray}
On the other hand, once the hyperons are present the following two
chemical imbalances need to be introduced in addition to
$\eta_{npe}$ and $\eta_{np\mu}$~\cite{Fernandez:2005cg}
\begin{eqnarray}
\eta_{2n\Sigma p}&=&2\mu_n-\mu_{\Sigma^-}-\mu_p,\\
\eta_{n\Lambda}&=&\mu_n-\mu_{\Lambda^0}
\end{eqnarray}
Since processes (19) and (20) do not conserve strangeness, they can
proceed only via weak interactions, while reactions (21) proceed via
strong interactions. The above reactions have timescales at least
several orders of magnitude shorter than those of beta
processes~\cite{LC1969} and, therefore, $\eta_{2n\Sigma p}$ and
$\eta_{n\Lambda}$ remain relatively small compared to $\eta_{npe}$
and $\eta_{np\mu}$~\cite{Fernandez:2005cg}. Consequently, reactions
(19)-(21) contribute negligibly to the total heat
generation~\cite{Fernandez:2005cg}.

The importance of including Urca processes involving hyperons in the
gravitochemical heating formalism, in addition to the nucleonic
processes, can be assessed, for instance, by considering the
following direct $\Sigma^-$ Urca reactions~\cite{Fernandez:2005cg}
\begin{eqnarray}
\Sigma^-&\rightarrow&n+l+\bar{\nu},\\
n+l&\rightarrow&\Sigma^-+\nu
\end{eqnarray}
The net effect of these processes on Eqs.~(8) and (9) is to enhance
the lepton direct Urca rates, reducing the chemical imbalances
($\eta_{npe},\eta_{np\mu}$) and, in turn, the surface temperature.
The overall correction is however small, because the direct Urca
reactions with hyperons are at least a factor 5 weaker than the
nucleon ones (see e.g. Prakash et al.~\cite{PPLP1992}). For the case
when only the modified Urca channel operates, Fern\'{a}ndez et
al.~\cite{Fernandez:2005cg} showed that the correction to the
surface temperature due to including several reactions involving
hyperons is on the order of
$[\tilde{L}_N/(\tilde{L}_N+\tilde{L}_H)]^{1/28}$, where
$\tilde{L}_N$ and $\tilde{L}_H$ are the nucleon and hyperon Urca
luminosities respectively. (This conclusion follows directly from
Eq.~(13).) Consequently, corrections due to hyperons in either
direct or modified Urca processes can be
neglected~\cite{Fernandez:2005cg}.

Yet, the inclusion of hyperons has a non-negligible effect on the
EOS of neutron-star matter: Due to the appearance of additional
degrees of freedom, the energy per particle is greatly reduced and
the resulting EOS is much softer than its purely nucleonic
counterpart~\cite{BBS2000,GFB2007} (see also Fig.~1). This effect is
entirely due to the inclusion of hyperons as additional degrees of
freedom and is observed in all current calculations independent of
the adopted many-body approach and/or interaction (see
e.g.~\cite{GFB2007}). The consequences for the neutron star global
properties, namely radii and masses, are reflected in an overall
very large mass reduction with respect to models of $npe\mu$-stars
(Fig.~3).

\begin{figure}[!t]
\centering
\includegraphics[totalheight=2.1in]{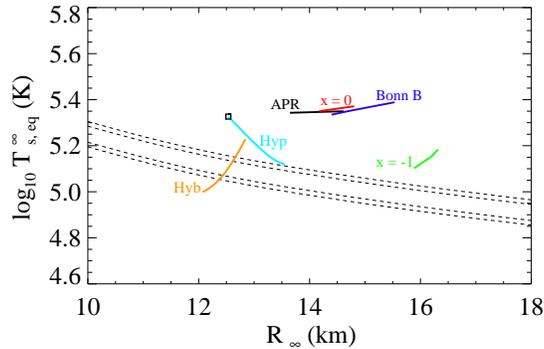}
\vspace{2mm} \caption{(Color online) Same as Fig.~5 but assuming
$|\dot{G}/G|=3.6\times 10^{-11}yr^{-1}$. The value of $\dot{G}$ is
chosen so that predictions from the hyperonic EOS (Hyp) are just
above the empirical constraints.}
\end{figure}

Thus, in light of the above considerations, to investigate the
impact of hyperons on the changing rate of $G$ we apply a hyperonic
EOS (see Section III) and the gravitochemical heating formalism
without including the imbalances (24) and (25) (i.e., we do not
consider openings of the hyperonic channels for additional heat
emission). Rather, we aim to identify the effect of the global
neutron star properties (masses and radii) from hyperonic EOS on the
time variations of $G$ in a qualitative way. In calculating the
surface temperature we adopt a fast cooling scenario since including
of hyperonic degrees of freedom increases proton fraction above the
threshold for fast {\it nucleonic} Urca processes for neutron stars
in the considered mass range (see Fig.~2 upper and middle frames).
We show our predictions for $|\dot{G}/G|$ based on the hyperonic EOS
(Hyp) in Fig.~9. The value $|\dot{G}/G|=3.6\times 10^{-11}yr^{-1}$
reassures that results from the Hyp EOS are (just) above the
empirical contours and is comparable with the one predicted with the
$x=-1$ EOS (which also allows for fast Urca reactions). Here a few
comments are in place: First, we should mention that due to the
``softness'' of the Hyp EOS, the maximum mass attained by the
neutron star models in this case is much smaller than the
corresponding masses predicted by purely nucleonic EOSs (see Fig.~3
and Table~1). The maximum mass is about $1.25M_{\odot}$ and falls
closer to the lower limit of the allowed mass range for PSR
J0437-4715. In fact, if one adopts the constraint by van Straten et
al.~\cite{vanStraten:2001zk} ($M_{PSR}=1.58\pm 0.18M_{\odot}$),
$M_{max}$ would be well below the allowed range. Also, as pointed
out by Burgio et al.~\cite{GFB2007}, a maximum mass less than
$1.3M_{\odot}$ is in contradiction with the most precisely measured
value of the Hulse-Taylor pulsar mass, PSR 1913+16, which amounts to
$1.44M_{\odot}$; Second, hyperonic stars are much more compact than
$npe\mu$-stars. With the Hyp EOS the maximum mass is achieved at
$\rho_c\approx 1.48fm^{-3}$ (see Table~2). The rapid rise of the
central density with increasing/decreasing neutron star mass/radius
causes a steep growth in the predicted surface temperature (see
e.g., Fig.~9); Finally, in the context of the gravitochemical
heating of interest here, the only appreciable effect of hyperons is
that their presence could facilitate the fulfilment of the direct
Urca conditions, as in the case of reactions involving
$\Lambda^0$~\cite{PPLP1992}. Additionally, the rise of the proton
fraction due to the appearance of $\Sigma^-$ hyperons can shift the
threshold of the nucleon direct Urca process to lower
densities~\cite{Glendenning}. These considerations, together with
the Hyp EOS predictions in Fig.~9, support the findings of
Reisenegger~\cite{AR}, namely that the EOSs do not give very
different results for the surface temperature if one only compares
among those which either do or do not allow for direct Urca
reactions.

{\it Quarks.-} The existence of quark matter in the interior of
neutron stars is one of the major issues in the physics of these
compact astrophysical objects. Here we discuss how the appearance of
deconfined quarks would affect a possible time variation of $G$. For
our analysis we choose a hybrid EOS (Hyb) derived by the Catania
group (for details, see Section III and the references therein).

Quark matter is assumed to consist of deconfined $u$, $d$, and $s$
quarks, and a small fraction of electrons~\cite{MBBS04}. At energies
relevant to neutron stars, the $u$ and $d$ quarks are treated as
massless particles, while the $s$ quark is moderately relativistic
and has a non-negligible mass~\cite{YKGH2001}. The most important
processes among the constituents of the quark-matter system are the
direct Urca reactions with $u$ and $d$ quarks~\cite{YKGH2001}
\begin{eqnarray}
d&\rightarrow&u+e+\bar{\nu}_e,\\
u+e&\rightarrow&d+\nu_e
\end{eqnarray}
In addition, the following direct Urca reactions involving the $s$
quark are possible:
\begin{eqnarray}
s&\rightarrow&u+e+\bar{\nu}_e,\\
u+e&\rightarrow&s+\nu_e
\end{eqnarray}
These $\beta$-processes, however, do not conserve strangeness and,
generally, yield much smaller emissivity relative to reactions (28)
and (29)~\cite{YKGH2001}. The chemical equilibrium conditions imply
\begin{equation}
\mu_d=\mu_s=\mu_u+\mu_e
\end{equation}

\begin{figure}[!t]
\centering
\includegraphics[totalheight=3.5in]{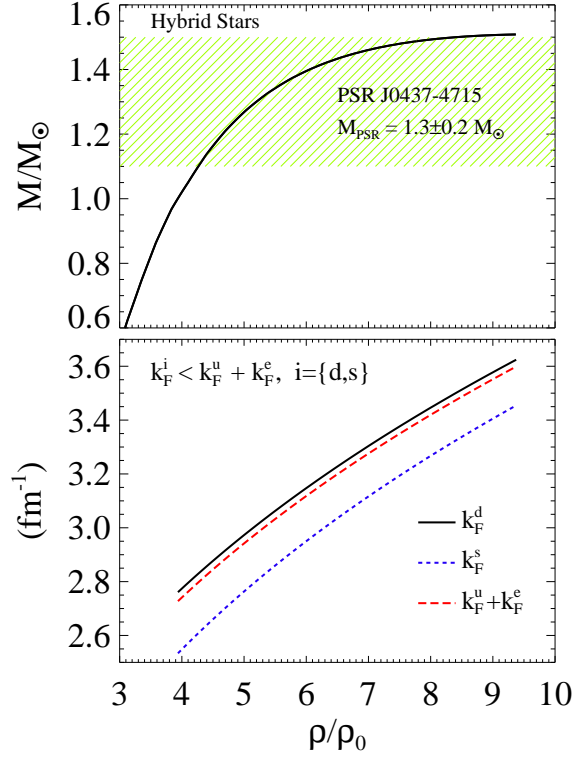}
\vspace{5mm} \caption{(Color online) Models of hybrid stars (upper
frame) and triangle inequality (lower frame).}
\end{figure}

\begin{figure}[!t]
\centering
\includegraphics[totalheight=2.1in]{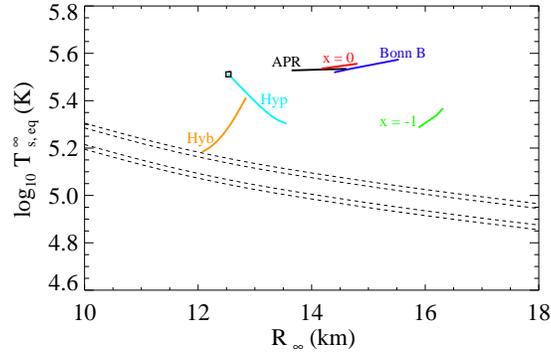}
\vspace{2mm} \caption{(Color online) Same as Fig.~5 but assuming
$|\dot{G}/G|=1.6\times 10^{-10}yr^{-1}$. The value of $\dot{G}$ is
chosen so that predictions from the hybrid EOS (Hyb) are just above
the empirical constraints.}
\end{figure}

A rigorous treatment of quark matter in the gravitochemical
formalism would require major adjustments/extensions of the
framework to incorporate, for instance, the imbalances $\eta_{ued}$
and $\eta_{ues}$, and the corresponding quark emissivities. Since
this task is beyond the main focus of this work, we take a much
simpler approach which should, at minimum, hint on the effect of
quarks relative to the $npe\mu$-matter. Similarly to what is done in
the case of hyperonic stars, we provide a (qualitative) estimate
based on the mass-radius correlation of hybrid stars, rather than
their detailed composition. Because the triangle inequality
\begin{equation}
k_F^i<k_F^u+k_F^e,\quad i=\{d,s\}
\end{equation}
among the Fermi momenta of species $u$, $d$, $s$ and $e$, is
fulfilled for reactions (30) and (31) but is not for (28) and (29)
(see Fig.~10), cooling of hybrid stars (from the Hyb EOS) can
proceed through quark direct Urca processes involving the $s$ quark.
Therefore we assume a fast-cooling scenario. Our predictions for
$|\dot{G}/G|$ based on the Hyb EOS are shown in Fig.~11. We notice
that the value $|\dot{G}/G|=1.6\times 10^{-10}yr^{-1}$ (chosen so
that predictions from the Hyb EOS are above the empirical data)
falls closest to the one derived by Jofre et al. \cite{Jofre:2006ug}
under the modified Urca assumption (and also our predictions from
the $x=-1$ and $Hyp$ EOSs). Since the yield of reactions (30) and
(31) is less certain \cite{YKGH2001} and because we do not treat the
inclusion of quarks explicitly, we should keep in mind that the
present result is more qualitative than quantitative and therefore
more refined calculations might be necessary to draw more definite
conclusions. On the other hand, we observe a steep rise of surface
temperature with stellar mass (red-shifted radius and/or central
density), see e.g. Fig.~11. The rapid temperature increase is
attributed to the fast growth of central density with mass (Fig.~10,
upper panel), and has been already observed in conjunction with the
analysis of hyperonic stars (see above). Finally, we notice that
hybrid stars have, generally, smaller radii than those of
$npe\mu$-stars (Fig.~3). In fact, the radii predicted with the Hyb
EOS employed here are below the latest neutron star radius
constraints~\cite{Li:2005sr} (see also Fig.~3). Yet, models yielding
quite stiff quark matter EOSs, and, in turn, hybrid/quark star radii
(masses) consistent with the latest observations~\cite{nature06}, do
exist~\cite{ABPR2005}. Given the large present uncertainties in
modeling the quark phase of matter, our results provide a reasonable
qualitative assessment of the changing rate of $G$, based on the
gravitochemical heating method and a hybrid EOS.

\section{Summary}

In summary, to test the Dirac hypothesis and constrain the changing
rate of the gravitational constant $G$ due to the continuous
expansion of the Universe, we have calculated the neutron star
surface temperature through the gravitochemical formalism introduced
by Jofr\'{e} et al.~\cite{Jofre:2006ug} applying several EOSs
spanning from pure nucleonic matter to quark matter. One of the
nucleonic EOSs has a density dependence of the symmetry energy
constrained by terrestrial nuclear laboratory data that just became
available recently. Using the ``softer'' symmetry energy ($x=0$)
consistent with both the isospin diffusion and the $^{208}Pb$
neutron skin data, we obtain an upper limit $|\dot{G}/G|\le4.5\times
10^{-12}yr^{-1}$ compatible with that obtained by Jofr\'{e} et
al.~\cite{Jofre:2006ug} under the same assumptions. This is mainly
because the density dependence of the symmetry energy with $x=0$
turns out to be  very close to that with the APR EOS they also used.
The ``stiffer'' symmetry energy ($x=-1$) EOS yields an upper limit
$|\dot{G}/G|\le2.1\times 10^{-11}yr^{-1}$, an order of magnitude
lower than the one derived by Jofr\'{e} et al.~\cite{Jofre:2006ug}
when both direct and modified Urca reactions operate. Since both the
$x=0$ and $x=-1$ equations of state have symmetry energies
consistent with the empirical nuclear data, we cannot rule out the
results from the direct Urca scenario. In the case of fast cooling,
our result provides even tighter upper limit than the one reported
in Ref.~\cite{Jofre:2006ug}.

The gravitochemical heating mechanism has the potential to become a
powerful tool for constraining gravitational physics. Since the
method relies on the detailed neutron star structure, which, in
turn, is determined by the EOS of stellar matter, further progress
in our understanding of properties of dense, neutron-rich matter
will make this approach more effective. Precise astrophysical
observations such as those by \"{O}zel~\cite{nature06}, Hessels et
al.~\cite{Hessels:2006ze}, and Kaaret at al.~\cite{Kaaret2007}
together with future heavy-ion experiments with high energy
radioactive beams~\cite{RIA} will allow us to set more stringent
constraints on the EOS of dense neutron-rich matter, and on the
possible time variation of the gravitational constant $G$.

\section*{Acknowledgements}

We would like to thank Rodrigo Fern\'{a}ndez and Andreas Reisenegger
for helpful discussions and assistance with the numerics. We also
thank Fiorella Burgio for providing the hyperonic and hybrid EOSs.
This work was supported by the National Science Foundation under
Grant No. PHY0652548 and the Research Corporation under Award No.
7123.

\end{document}